\documentclass[12pt]{article}
\usepackage[latin9]{inputenc}
\usepackage[active]{srcltx}
\usepackage{amsmath}
\usepackage{amssymb}
\usepackage[unicode=true,pdfusetitle,
 bookmarks=true,bookmarksnumbered=false,bookmarksopen=false,
 breaklinks=false,pdfborder={0 0 0},pdfborderstyle={},backref=false,colorlinks=false]
 {hyperref}

\makeatletter
\numberwithin{equation}{section}

\usepackage{amsfonts}
\usepackage[square,numbers,sort&compress]{natbib}
\makeatother

\begin{document}
\begin{center}
\textbf{\large{}Spinors, Proper Time and Higher-Spin Fields}{\large\par}
\par\end{center}

\begin{center}
\vspace{0.2cm}
 \textbf{N.G.~Misuna}\\
 \vspace{0.5cm}
 \emph{Max-Planck-Institut für Gravitationsphysik (Albert-Einstein-Institut),}\\
 \emph{ Am Mühlenberg 1, 14476, Potsdam, Germany }\\
 \vspace{0.5cm}
 \textit{Tamm Department of Theoretical Physics, Lebedev Physical
Institute,}\\
 \textit{Leninsky prospekt 53, 119991, Moscow, Russia}\\
 
\par\end{center}

\begin{center}
\vspace{0.6cm}
 nikita.misuna@aei.mpg.de \\
 
\par\end{center}

\vspace{0.4cm}

\begin{abstract}
\noindent We present a Lagrangian formulation for $4d$ integer-spin
relativistic fields in the $5d$ space spanned by two conjugate Weyl
spinors and a Lorentz-invariant proper-time coordinate. We construct
a manifestly Poincaré-invariant free classical action, find a general
solution to equations of motion and a corresponding positive-definite
inner product. Our formulation displays a separation of variables:
equations of motion represent ODE in a proper time only, while spinor
coordinates parameterize the Cauchy hypersurface. We also find momentum
eigenstates solutions for massless arbitrary integer-spin fields and
a massive scalar field.

\newpage{}

\tableofcontents{}
\end{abstract}

\section{Introduction}

Higher-spin (HS) theories represent an important class of models of
fundamental interactions. Covariant Lagrangian formulations for free
higher-spin fields have been constructed in massive case by Singh
and Hagen \cite{SinghHagen1,SinghHagen2}, and in massless case by
Fronsdal and Fang, both in Minkowski \cite{Fronsdal,FangFronsdal}
and (A)dS \cite{FronsdalCurved,FangFronsdalCurved} spaces. But it
turned out that constructing consistent interactions for massless
HS fields, which problem is of the most interest, gets very involved
in the covariant setup. Therefore the main progress beyond the free
level is due to other approaches.

In particular, cubic HS interactions have been found and studied in
detail within the light-cone framework (see e.g. \cite{lcone1,lcone2,lcone3,lcone4,lcone5}).
However, already beyond the cubic level the analysis becomes too complicated.

Self-dual HS models are conveniently formulated and analyzed by means
of the methods of twistor theory \cite{tw1,tw2,tw3,tw4}.

The full all-order system of classical e.o.m. of interacting HS gauge
fields has been constructed by Vasiliev \cite{vas1,vas2} in terms
of the generating equations, written in the so-called unfolded form
\cite{unf1,unf2,unf3} (for a review of Vasiliev theory see \cite{VasReview1,VasReview2}).
But extracting HS vertices from Vasiliev equations represents a very
nontrivial task, because one must restrict somehow the degree of non-locality
while solving for auxiliary generating variables, which problem is
currently under the active study (see \cite{local} and references
therein).

More references and a partial review of the recent HS literature can
be found in \cite{snowmass}.

Thus, the availability of different implementations of HS fields significantly
enriches our possibilities for constructing and studying HS theories.
In this paper we propose a new realization for the integer-spin representations
of the $4d$ Poincaré group. Instead of dealing with $4d$ Minkowski
space, we consider a $5d$ space spanned by a pair of conjugate spinors
and one Lorentz scalar. This set of coordinates appeared previously
in the unfolded formulation of the $4d$ off-shell fields \cite{misuna1,misuna2,misuna3,misuna4},
where they have been playing the role of the auxiliary fiber coordinates,
encoding unfolded descendants of the space-time fields under consideration.
In this paper we use these coordinates to build a self-contained Lagrangian
formulation for $4d$ integer-spin fields without any reference to
a space-time.

To give a preliminary intuitive idea of how such $5d$ space can encode
$4d$ fields, let us consider a simple example. An asymptotic one-particle
state of a scalar field is determined by 4-momentum $p^{a}=(E,\overrightarrow{p})$,
which is forced to lie on the mass-shell $p_{a}p^{a}=m^{2}$. Hence,
the state is fixed by three independent parameters: four variables
with one constraint. Alternatively, the same information can be encoded
in a Lorentz-scalar $\pi=\sqrt{E^{2}-\overrightarrow{p}^{2}}$ and
a null vector $n^{a}=(|\overrightarrow{p}|,\overrightarrow{p})$,
with the constraint being $\pi=m$. In its turn, a real null $4d$
vector can be represented in terms of spinors as $n^{a}=(\bar{\sigma})^{\dot{\alpha}\beta}\bar{\xi}_{\dot{\alpha}}\xi_{\beta}$.
Thus, a set of 5 variables $\{\pi,\xi^{\alpha},\bar{\xi}^{\dot{\alpha}}\}$
(effectively, 4 of them, as the global phase of $\xi$ does not contribute)
determines the 4-momentum, while the mass-shell equation becomes simply
$\pi=m$, putting no restrictions on $\xi$.

In our consideration, however, we make use of a similar $5d$ space
as a substitute not for the momentum $p^{a}$, but rather for the
coordinate $x^{a}$, so that classical e.o.m. become ODE in a scalar
coordinate. We find expressions for Poincaré generators and identify
appropriate modules supplied with a positive-definite inner product.We
also construct simple Poincaré-invariant actions which lead to the
appropriate e.o.m. and find their general solutions. In addition,
we find solutions for momentum eigenstates for the cases of an arbitrary-mass
scalar field and of massless arbitrary spin fields.

The paper is organized as follows. In Section \ref{SEC_Poincare_fields}
we introduce our conventions for Poincaré generators and give a brief
reminder on how covariant quantum fields are constructed in the standard
approach, to be later compared with our construction. In Section \ref{SEC_spin_rep}
we build a $4d$ integer-spin representation on a certain $5d$ space.
In Section \ref{SEC_action_eom} we present a Poincaré-invariant action
for a free field, give a general solution to e.o.m. and propose an
inner product for solutions. In Section \ref{SEC_plane_waves} we
find solutions of e.o.m. corresponding to momentum eigenstates for
a scalar field and massless fields. In Section \ref{SEC_conclusions}
we sum up our results.

\section{$4d$ Poincaré algebra and relativistic fields\label{SEC_Poincare_fields}}

Elementary particles are associated with unitary irreducible representations
(UIRs) of the Poincaré group (or an isometry group of the spacetime
in question, more generally) \cite{Wigner}.

In the paper we consider $4d$ Poincaré algebra with generators $P_{\alpha\dot{\alpha}}$,
$M_{\alpha\beta}=M_{\beta\alpha}$ and $\bar{M}_{\dot{\alpha}\dot{\beta}}=\bar{M}_{\dot{\beta}\dot{\alpha}}$,
which correspond to translations, anti-selfdual and selfdual rotations
of Minkowski space, respectively. Here indices belong to two conjugate
spinor representations of the Lorentz algebra $sl(2,\mathbb{C})$.
Commutation relations are
\begin{align}
 & [M_{\alpha\beta},M_{\gamma\delta}]=\epsilon_{\alpha\gamma}M_{\beta\delta}+\epsilon_{\alpha\delta}M_{\beta\gamma}+\epsilon_{\beta\gamma}M_{\alpha\delta}+\epsilon_{\beta\delta}M_{\alpha\gamma},\label{=00005BM,M=00005D}\\
 & [\bar{M}_{\dot{\alpha}\dot{\beta}},\bar{M}_{\dot{\gamma}\dot{\delta}}]=\epsilon_{\dot{\alpha}\dot{\gamma}}\bar{M}_{\dot{\beta}\dot{\delta}}+\epsilon_{\dot{\alpha}\dot{\delta}}\bar{M}_{\dot{\beta}\dot{\gamma}}+\epsilon_{\dot{\beta}\dot{\gamma}}\bar{M}_{\dot{\alpha}\dot{\delta}}+\epsilon_{\dot{\beta}\dot{\delta}}\bar{M}_{\dot{\alpha}\dot{\gamma}},\label{=00005BMbar,Mbar=00005D}\\
 & [M_{\alpha\beta},\bar{M}_{\dot{\gamma}\dot{\delta}}]=0,\label{=00005BM,Mbar=00005D}\\
 & [M_{\alpha\beta},P_{\gamma\dot{\gamma}}]=\epsilon_{\alpha\gamma}P_{\beta\dot{\gamma}}+\epsilon_{\beta\gamma}P_{\alpha\dot{\gamma}},\label{=00005BM,P=00005D}\\
 & [\bar{M}_{\dot{\alpha}\dot{\beta}},P_{\gamma\dot{\gamma}}]=\epsilon_{\dot{\alpha}\dot{\gamma}}P_{\gamma\dot{\beta}}+\epsilon_{\dot{\beta}\dot{\gamma}}P_{\gamma\dot{\alpha}},\label{=00005BMbar,P=00005D}\\
 & [P_{\alpha\dot{\alpha}},P_{\beta\dot{\beta}}]=0,\label{=00005BP,P=00005D}
\end{align}
where $\epsilon_{\alpha\beta}$ and $\epsilon_{\dot{\alpha}\dot{\beta}}$
are Lorentz-invariant spinor metrics
\begin{equation}
\epsilon_{\alpha\beta}=\epsilon^{\alpha\beta}=\epsilon_{\dot{\alpha}\dot{\beta}}=\epsilon^{\dot{\alpha}\dot{\beta}}=\left(\begin{array}{cc}
0 & 1\\
-1 & 0
\end{array}\right),\label{epsilon_tensor}
\end{equation}
which raise and lower spinor indices according to
\begin{equation}
v_{\alpha}=\epsilon_{\beta\alpha}v^{\beta},\quad v^{\alpha}=\epsilon^{\alpha\beta}v_{\beta},\quad\bar{v}_{\dot{\alpha}}=\epsilon_{\dot{\beta}\dot{\alpha}}\bar{v}^{\dot{\beta}},\quad\bar{v}^{\dot{\alpha}}=\epsilon^{\dot{\alpha}\dot{\beta}}\bar{v}_{\dot{\beta}}.\label{index_moves}
\end{equation}

UIRs are determined by the values of two Casimir operators: a square
of the momentum, associated with the mass,
\begin{equation}
P^{2}=m^{2}\label{P^2=00003Dm^2}
\end{equation}
and, introducing the Pauli\textendash Lubanski pseudovector as
\begin{equation}
W_{\alpha\dot{\alpha}}=\frac{1}{2}M_{\alpha\beta}P^{\beta}\text{}_{\dot{\alpha}}-\frac{1}{2}\bar{M}_{\dot{\alpha}\dot{\beta}}P_{\alpha}\text{}^{\dot{\beta}},\label{W_def}
\end{equation}
either its square, associated with the spin $s$ when $m^{2}>0$
\begin{equation}
W^{2}=-m^{2}s(s+1),\label{W^2=00003Dm^2s(s+1)}
\end{equation}
or the helicity $\lambda$ when $m=0$
\begin{equation}
W_{\alpha\dot{\beta}}=\lambda P_{\alpha\dot{\beta}}.\label{W=00003Dlambda_P}
\end{equation}
In \eqref{P^2=00003Dm^2}, \eqref{W^2=00003Dm^2s(s+1)} and throughout
the paper the square $v^{2}$ of a vector $v_{\alpha\dot{\beta}}$
is defined as
\begin{equation}
v^{2}=\frac{1}{2}v_{\alpha\dot{\beta}}v^{\alpha\dot{\beta}}.
\end{equation}

The standard covariant QFT approach is to implement momentum generators
as coordinate derivatives
\begin{equation}
P_{a}=-i\frac{\partial}{\partial x^{a}}\label{P_standard}
\end{equation}
on the Minkowski space with coordinates $x^{a}$. Then quantum fields
look as $\phi^{I}(x)$, where index $I$ belongs to some finite-dimensional
representation of the Lorentz group (spin), so that rotations are
realized as
\begin{equation}
M_{a,b}=i(x_{a}\frac{\partial}{\partial x^{b}}-x_{b}\frac{\partial}{\partial x^{a}})+(S_{a,b})^{I}{}_{J}\label{M_standard}
\end{equation}
with $S$ being $x$-independent spin generators. In general, however,
the resulting representation of the Poincaré algebra is neither irreducible
nor unitary, and one has to remove undesirable subrepresentations
by imposing additional constraints besides the Klein\textendash Gordon
equation \eqref{P^2=00003Dm^2}. In order to represent all of them
as following from some Lagrangian equations of motion, one has to
introduce auxiliary fields (for massive fields with $s>1$) and/or
to provide certain gauge symmetry (for massless fields with $s\geq1$).
Corresponding  Lagrangian formulations for arbitrary spin fields have
been constructed by Sing and Hagen for massive fields \cite{SinghHagen1,SinghHagen2}
and by Fronsdal and Fang for massless fields \cite{Fronsdal,FangFronsdal,FronsdalCurved,FangFronsdalCurved}.

\section{Spin-$s$ representation\label{SEC_spin_rep}}

In the paper we construct a realization of bosonic UIRs on a $5d$
linear space spanned by a pair of conjugate commuting $sl(2,\mathbb{C})$
spinors $Y^{A}=(y^{\alpha},\bar{y}^{\dot{\alpha}})$ and a Lorentz-invariant
'proper time' $\tau$. This set of variables $(Y,\tau)$ was previously
used in formulating off-shell unfolded equations for various $4d$
field systems \cite{misuna1,misuna2,misuna3,misuna4}. And spinors
$Y$ were initially used in the unfolded Vasiliev equations \cite{vas1,vas2},
where they play the crucial role of the generators of an associative
HS gauge algebra. Here we propose to use $(Y,\tau)$-space instead
of a space-time and build a corresponding Lagrangian formulation for
bosonic fields. All fields are 'scalar' (i.e. without non-contracted
Lorentz indices) functions $F(Y,\tau)$ on this space.

For the rotation generators we take
\begin{equation}
M_{\alpha\beta}=y_{\alpha}\partial_{\beta}+y_{\beta}\partial_{\alpha},\label{M_def}
\end{equation}
\begin{equation}
\bar{M}_{\dot{\alpha}\dot{\beta}}=\bar{y}_{\dot{\alpha}}\bar{\partial}_{\dot{\beta}}+\bar{y}_{\dot{\beta}}\bar{\partial}_{\dot{\alpha}},\label{Mbar_def}
\end{equation}
where $Y$-derivatives are defined as
\begin{equation}
\partial_{\alpha}y^{\beta}=\delta_{\alpha}\text{}^{\beta},\quad\bar{\partial}_{\dot{\alpha}}\bar{y}^{\dot{\beta}}=\delta_{\dot{\alpha}}\text{}^{\dot{\beta}}.\label{dy=00003D1}
\end{equation}
It is easy to check that \eqref{M_def}-\eqref{Mbar_def} satisfy
\eqref{=00005BM,M=00005D}-\eqref{=00005BM,Mbar=00005D}. From here
it also directly follows that the proper-time coordinate $\tau$ is
Lorentz-invariant (but not translation-invariant, as we will see).
The expressions \eqref{M_def}-\eqref{Mbar_def} for rotations operators
are universal: we demand that they look the same for all fields of
arbitrary masses and spins, like it is the case for the translation
operator in the standard construction \eqref{P_standard}. The price
to pay for this is that the translation operator now depends on a
spin, as we will see.

As $Y$ commute with themselves, they have zero norm
\begin{equation}
y_{\alpha}y^{\alpha}=0,\quad\bar{y}_{\dot{\alpha}}\bar{y}^{\dot{\alpha}}=0,\label{y^2=00003D0}
\end{equation}
and the only independent Lorentz-invariant $Y$-combinations one can
form are Euler operators
\begin{equation}
N=y^{\alpha}\partial_{\alpha},\quad\bar{N}=\bar{y}^{\dot{\alpha}}\bar{\partial}_{\dot{\alpha}}.\label{N_Nbar_def}
\end{equation}

An appropriate module of a spin-$s$ representation has to contain
states with helicities from $-s$ to $+s$. This can be achieved by
considering a set of functions 
\begin{equation}
\Phi^{s}(Y,\tau)=\{\Phi_{\alpha(m),\dot{\alpha}(n)}(\tau)(y^{\alpha})^{m}(\bar{y}^{\dot{\alpha}})^{n},\quad(m+n)\geq2s,\quad|m-n|\leq2s\},\label{Phi_set_module}
\end{equation}
where we make use of condensed notations for symmetrized indices
\begin{equation}
v_{\alpha(m)}=v_{(\alpha_{1}\alpha_{2}...\alpha_{m})},\quad(y^{\alpha})^{m}=y^{\alpha_{1}}y^{\alpha_{2}}...y^{\alpha_{m}}.\label{condens_notat}
\end{equation}
The module \eqref{Phi_set_module} can be also represented as
\begin{equation}
\Phi^{s}(Y,\tau)=\Phi_{A(2s)}(y\bar{y},\tau)(Y^{A})^{2s},\label{Phi_module}
\end{equation}
where $A$ is a Majorana index taking four values $\{1,2,\dot{1},\dot{2}\}$.
This form is visually more similar to the standard Minkowski approach,
where an integer spin-$s$ module is a rank-$s$ tensor field $\phi_{a(s)}(x)$.
It should be stressed however, that in \eqref{Phi_module} 'external'
$Y$-s and 'internal' $y$-s and $\bar{y}$-s are on a completely
equal footing, as seen from \eqref{Phi_set_module}. And $2s$ explicit
spinors and indices in \eqref{Phi_module} are highlighted only in
order to show restrictions on the number of $y$ and $\bar{y}$ and
play no special role otherwise.

Now one has to find an expression for the momentum operator $P_{\alpha\dot{\beta}}$.
The most general Ansatz is
\begin{equation}
P_{\alpha\dot{\beta}}=a_{N,\bar{N}}\partial_{\alpha}\bar{\partial}_{\dot{\beta}}+b_{N,\bar{N}}y_{\alpha}\bar{y}_{\dot{\beta}}+c_{N,\bar{N}}y_{\alpha}\bar{\partial}_{\dot{\beta}}+\bar{c}_{N,\bar{N}}\partial_{\alpha}\bar{y}_{\dot{\beta}},\label{P_Ansatz}
\end{equation}
where Lorentz-invariant coefficients $a,b,c,\bar{c}$ are built out
of Euler operators \eqref{N_Nbar_def}, as well as of $\tau$ and
$\tau$-derivatives. \eqref{P_Ansatz} automatically satisfies \eqref{=00005BM,P=00005D}
and \eqref{=00005BMbar,P=00005D}, so the only equation to be solved
is \eqref{=00005BP,P=00005D}. It can be equivalently reformulated
in terms of two conjugate equations 
\begin{equation}
P_{\alpha\dot{\beta}}P_{\alpha\dot{\gamma}}\epsilon^{\dot{\beta}\dot{\gamma}}=0,\label{PP_0_1}
\end{equation}
\begin{equation}
P_{\beta\dot{\alpha}}P_{\gamma\dot{\alpha}}\epsilon^{\beta\gamma}=0.\label{PP_0_2}
\end{equation}
Substituting \eqref{P_Ansatz}, they lead to the following constraints
\begin{eqnarray}
 &  & (\bar{N}+2)a_{N,\bar{N}}\bar{c}_{N+1,\bar{N}+1}-\bar{N}a_{N+1,\bar{N}-1}\bar{c}_{N,\bar{N}}=0,\label{cons1}\\
 &  & (\bar{N}+2)b_{N-1,\bar{N}+1}c_{N,\bar{N}}-\bar{N}b_{N,\bar{N}}\bar{c}_{N-1,\bar{N}-1}=0,\label{cons2}\\
 &  & (\bar{N}+2)a_{N,\bar{N}}b_{N+1,\bar{N}+1}-\bar{N}a_{N-1,\bar{N}-1}b_{N,\bar{N}}+(\bar{N}+2)c_{N,\bar{N}}\bar{c}_{N-1,\bar{N}+1}-\bar{N}\bar{c}_{N,\bar{N}}c_{N+1,\bar{N}-1}=0,\nonumber \\
\label{cons3}
\end{eqnarray}
plus three conjugate equations with $N\leftrightarrow\bar{N}$, $c\leftrightarrow\bar{c}$
interchanged. In addition, one has to ensure that the action of \eqref{P_Ansatz}
does not lead outside the module \eqref{Phi_set_module}. This means
that only those solutions are suitable that satisfy
\begin{equation}
a_{N,\bar{N}}|_{\varsigma=s-1}=0,\quad c_{N,\bar{N}}|_{\chi=s+1}=0,\quad\bar{c}_{N,\bar{N}}|_{\chi=-s-1}=0,\label{bndr_cond}
\end{equation}
where $\varsigma$ and $\chi$ are important linear combinations of
Euler operators \eqref{N_Nbar_def}, which we actively use below,
\begin{equation}
\varsigma=\frac{N+\bar{N}}{2},\quad\chi=\frac{N-\bar{N}}{2}.\label{sigma_chi_def}
\end{equation}
Any solution of \eqref{cons1}-\eqref{cons3} respecting boundary
conditions \eqref{bndr_cond} defines some representation of the Poincaré
algebra. But many of these representations are equivalent, and this
allows one to put some further constraints.

First, we restrict $\tau$-dependence and provide a 'separation of
variables' $Y$ and $\tau$. Specifically, we require the operator
$P^{2}$ to be $Y$-independent, so that the mass-shell equation \eqref{P^2=00003Dm^2}
becomes an ODE in $\tau$. In addition, we demand $\tau$ to enter
\eqref{P_Ansatz} only through this $P^{2}$-combination.

Second, we require $P_{\alpha\dot{\beta}}$ to allow for a usual integration
by parts rule
\begin{equation}
\int d\tau\int d^{4}Yf(Y,\tau)P_{\alpha\dot{\beta}}g(Y,\tau)=-\int d\tau\int d^{4}Yg(Y,\tau)P_{\alpha\dot{\beta}}f(Y,\tau).\label{part_integr}
\end{equation}
To this end one notes that (assuming that one can neglect boundary
terms)
\begin{equation}
\int d^{4}Y(y^{\alpha}\partial_{\alpha}f(Y))g(Y)=\int d^{4}Y((\partial_{\alpha}y^{\alpha}-2)f(Y))g(Y)=-\int d^{4}Yf(Y)(y^{\alpha}\partial_{\alpha}+2)g(Y),
\end{equation}
which allows one to formulate general rules
\begin{equation}
\int Nf\cdot g=-\int f\cdot(N+2)g,\quad\int\bar{N}f\cdot g=-\int f\cdot(\bar{N}+2)g,\quad\int\varsigma f\cdot g=-\int f\cdot(\varsigma+2)g,\quad\int\chi f\cdot g=-\int f\cdot\chi g.
\end{equation}
These constraints significantly restrict the space of solutions to
\eqref{PP_0_1}-\eqref{PP_0_2}, though still do not fix it unambiguously.
We pick up the following particular solution
\begin{eqnarray}
-iP_{\alpha\dot{\beta}} & = & \frac{(\varsigma-s+1)(\varsigma+s+2)(\varsigma+3/2)}{(N+1)(N+2)(\bar{N}+1)(\bar{N}+2)}\partial_{\alpha}\bar{\partial}_{\dot{\beta}}-\frac{P^{2}}{(\varsigma+1/2)}y_{\alpha}\bar{y}_{\dot{\beta}}+\nonumber \\
 &  & +\frac{1}{(\bar{N}+1)(\bar{N}+2)}[(\chi+s)(\chi-s-1)\Pi^{+}-P^{2}\Pi^{-0}]y_{\alpha}\bar{\partial}_{\dot{\beta}}+\nonumber \\
 &  & +\frac{1}{(N+1)(N+2)}[(\chi-s)(\chi+s+1)\Pi^{-}-P^{2}\Pi^{+0}]\partial_{\alpha}\bar{y}_{\dot{\beta}},\label{P_solution}
\end{eqnarray}
where projectors $\Pi$ on different $\chi$-components are introduced
as

\begin{eqnarray}
 &  & \Pi^{+}F_{\chi}(Y)=\begin{cases}
F_{\chi}(Y), & \chi>0\\
0, & \chi\leq0
\end{cases};\qquad\Pi^{-}F_{\chi}(Y)=\begin{cases}
F_{\chi}(Y), & \chi<0\\
0, & \chi\geq0
\end{cases};\\
 &  & \Pi^{+0}F_{\chi}(Y)=\begin{cases}
F_{\chi}(Y), & \chi\geq0\\
0, & \chi<0
\end{cases};\qquad\Pi^{-0}F_{\chi}(Y)=\begin{cases}
F_{\chi}(Y), & \chi\leq0\\
0, & \chi>0
\end{cases}.
\end{eqnarray}
Expression \eqref{P_solution} for $P$ contains manifestly and self-consistently
its own square $P^{2}$, which is $Y$-independent by construction.
$P^{2}$ is also required to be even under integration by parts in
order to provide \eqref{part_integr}.

Now for the Pauli\textendash Lubanski pseudovector \eqref{W_def}
one has
\begin{eqnarray}
-iW_{\alpha\dot{\beta}} & = & -\chi\frac{(\varsigma-s+1)(\varsigma+s+2)(\varsigma+3/2)}{(N+1)(N+2)(\bar{N}+1)(\bar{N}+2)}\partial_{\alpha}\bar{\partial}_{\dot{\beta}}-\chi\frac{P^{2}}{(\varsigma+1/2)}y_{\alpha}\bar{y}_{\dot{\beta}}+\nonumber \\
 &  & +\frac{(\varsigma+1)}{(\bar{N}+1)(\bar{N}+2)}[(\chi+s)(\chi-s-1)\Pi^{+}-P^{2}\Pi^{-0}]y_{\alpha}\bar{\partial}_{\dot{\beta}}-\nonumber \\
 &  & -\frac{(\varsigma+1)}{(N+1)(N+2)}[(\chi-s)(\chi+s+1)\Pi^{-}-P^{2}\Pi^{+0}]\partial_{\alpha}\bar{y}_{\dot{\beta}},\label{W_solution}
\end{eqnarray}
with its square being
\begin{equation}
W^{2}=-P^{2}s(s+1).
\end{equation}
In the case $P^{2}=0$ one finds that $P_{\alpha\dot{\beta}}$ and
$W_{\alpha\dot{\beta}}$ are proportional to each other whenever the
module contains components with $|\chi|=s$ only, in which case
\begin{eqnarray}
-iP_{\alpha\dot{\beta}}^{m=0} & = & \frac{(\varsigma-s+1)(\varsigma+s+2)(\varsigma+3/2)}{(N+1)(N+2)(\bar{N}+1)(\bar{N}+2)}\partial_{\alpha}\bar{\partial}_{\dot{\beta}},\quad W_{\alpha\dot{\beta}}^{m=0}=-\chi P_{\alpha\dot{\beta}}^{m=0},
\end{eqnarray}
that corresponds to two $\pm s$ helicities \eqref{W=00003Dlambda_P}
of the massless field.

Thus, operators \eqref{M_def}, \eqref{Mbar_def} and \eqref{P_solution}
indeed correctly determine a spin-$s$ representation on the module
\eqref{Phi_set_module} after fixing the value of $P^{2}$. In the
massless case $P^{2}=0$ one also has to reduce the module, leaving
only $\pm s$ helicities, which corresponds to setting $|m-n|=2s$
instead of $|m-n|\leq2s$ in \eqref{Phi_set_module}, or having, instead
of \eqref{Phi_module},
\begin{equation}
\Phi_{m=0}^{s}(Y,\tau)=\Phi_{\alpha(2s)}(y\bar{y},\tau)(y^{\alpha})^{2s}\oplus\bar{\Phi}_{\dot{\alpha}(2s)}(y\bar{y},\tau)(\bar{y}^{\dot{\alpha}})^{2s}.\label{Phi_module_massless}
\end{equation}
Now, in order to formulate an action principle, one has to realize
$P^{2}$ as a differential operator. As mentioned previously, it must
be $\tau$-dependent only and even under integration by parts, but
completely unrestricted otherwise. This means that in our construction
Klein\textendash Gordon equation \eqref{P^2=00003Dm^2} can be implemented
in many different ways. In the next Section we consider one of the
simplest possibilities.

\section{Free action, e.o.m. and inner product\label{SEC_action_eom}}

First we consider the massive case. We take
\begin{equation}
P^{2}=-\frac{\partial^{2}}{\partial\tau^{2}}.\label{P^2=00003Dd^2}
\end{equation}
Then a Poincaré-invariant action for a spin-$s$ mass-$m$ field is
simply
\begin{equation}
S=\frac{1}{2}\sum_{\chi=-s}^{s}\int d^{4}Y\int d\tau(\dot{\Phi}_{\chi}^{2}-m^{2}\Phi_{\chi}^{2}),\label{S_action}
\end{equation}
where the dot means a $\tau$-derivative and $\Phi_{\chi}$ means
a subspace of the spin-$s$ module \eqref{Phi_set_module} of the
definite helicity-$\chi$
\begin{equation}
\Phi_{\chi}(Y,\tau)=\Phi_{\alpha(s+\chi),\dot{\beta}(s-\chi)}(y\bar{y},\tau)(y^{\alpha})^{s+\chi}(y^{\dot{\beta}})^{s-\chi}.\label{Phi_chi}
\end{equation}
Poincaré-invariance of the action \eqref{S_action} is guaranteed
by the integration-by-parts property \eqref{part_integr}, which is
obvious for $M$ and $\bar{M}$ \eqref{M_def}-\eqref{Mbar_def} as
well.

The action \eqref{S_action} leads to an e.o.m.
\begin{equation}
\ddot{\Phi}_{\chi}+m^{2}\Phi_{\chi}=0.\label{eom}
\end{equation}
Its general solution is
\begin{equation}
\Phi_{\chi}(Y,\tau)=e^{-im\tau}f_{\chi}(Y)+e^{im\tau}g_{\chi}(Y),\label{gen_sol}
\end{equation}
where the only requirement to $Y$-functions $f$ and $g$ is to belong
to helicity-$\chi$ subspace. Thus, from the point of view of \eqref{eom},
$Y$ are coordinates on the subspace of Cauchy data, while e.o.m.
determines the evolution in $\tau$-direction. 

A Poincaré-invariant inner product for the on-shell states is
\begin{equation}
(\text{\ensuremath{\Phi}}_{\chi},\Psi_{\chi'})=i\int d^{4}Y(\bar{\ensuremath{\Phi}}\dot{\Psi}-\Psi\dot{\bar{\ensuremath{\Phi}}})\delta_{\chi,\chi'}.\label{inner_product}
\end{equation}
It is $\tau$-independent due to \eqref{eom} and positive-definite
for a 'positive-mass' subspace of \eqref{gen_sol} with $g=0$. The
states with the same $Y$-dependence but with different mass signs
are orthogonal.

The split of the on-shell space into two subspaces, corresponding
to 'positive-mass' $f$ and 'negative-mass' $g$ contributions in
\eqref{inner_product}, is reminiscent to the split into positive-energy
and negative-energy branches in the standard QFT. However, establishing
the rigorous relation between two these phenomenae requires a separate
thorough analysis which we leave for the future study. Let us note,
however, that in our case the split, being determined by $\tau$-dependence,
is manifestly Lorentz-invariant.

Now we move to the massless case. Here using \eqref{P^2=00003Dd^2}
potentially leads to problems: the general solution to \eqref{eom}
with $m=0$ is an arbitrary linear function of $\tau$, so all on-shell
states either have zero norm with respect to \eqref{inner_product}
or are unbounded in $\tau$, which may be unpleasant.

This can be easily fixed by introducing a mass-dimension parameter
$\mu$ and deforming \eqref{P^2=00003Dd^2} to
\begin{equation}
P^{2}=-\frac{\partial^{2}}{\partial\tau^{2}}-\mu^{2}.\label{P^2_mu}
\end{equation}
Then the zero-mass action becomes
\begin{equation}
S=\frac{1}{2}\sum_{\chi=-s}^{s}\int d^{4}Y\int d\tau(\dot{\Phi}_{\chi}^{2}-\mu^{2}\Phi_{\chi}^{2}),\label{S_action-1}
\end{equation}
and e.o.m. now are
\begin{equation}
\ddot{\Phi}_{\chi}+\mu^{2}\Phi_{\chi}=0,\label{eom_mu}
\end{equation}
so the general solution is
\begin{equation}
\Phi_{\chi}(Y,\tau)=e^{-i\mu\tau}f_{\chi}(Y)+e^{i\mu\tau}g_{\chi}(Y),\label{gen_sol_massless}
\end{equation}
and one has $\tau$-bounded functions and the split into two branches
again.

As said before, in the massless case one also has to reduce the module,
leaving only $|\chi|=s$ components, \eqref{Phi_module_massless}.
Intermediate components $|\chi|<s$ are necessary to provide off-shell
Poincaré invariance of the action \eqref{S_action-1}, but on shell
$|\chi|=s$ components decouple into closed subspaces.

It should be stressed that the equation \eqref{eom_mu} describes
a massless field, $m=0$. The parameter $\mu$ does not shift the
value of the mass, it only deforms the functional dependence of $P^{2}$
on $\tau$. In particular, $\mu$ enters directly the expression for
the off-shell momentum generator \eqref{P_solution} through \eqref{P^2_mu}.
In principle, it can be introduced for the massive fields as well.
Practically, the parameter $\mu$ plays the role of a manifestly Poincaré-invariant
IR-regulator. The possibility of such deformation relies on the large
freedom in choosing the differential realization of the $P^{2}$ and
is specific to the presented construction. In particular, it is unclear
how to locally deform the momentum operator \eqref{P_standard} of
a covariant QFT to have $P^{2}=-\square+\mu^{2}$.

Let us also give a brief comment on the issue of locality of the constructed
representations. As seen from \eqref{P_solution}, the translations,
as opposite to the rotations \eqref{M_def}-\eqref{Mbar_def}, are
realized non-locally: $Y$-differential operators $N$ and $\bar{N}$
enter \eqref{P_solution} in a non-polynomial way. But a crucial feature
is that the translations are local in $\tau$, so one cannot e.g.
shift the pole of the propagator by means of Poincaré-transformations.
So the evolution in $\tau$ is completely local, while transformations
on the Cauchy hypersurface with coordinates $Y$ are non-local. 

\section{Momentum eigenstates\label{SEC_plane_waves}}

Having formulated the classical action and e.o.m., the next natural
step is to look for various partial solutions to them. Of special
importance are solutions that correspond to momentum eigenstates.
We restrict ourselves here to the simplest cases of a scalar field
and massless arbitrary spin fields, for which the momentum operator
takes a particularly simple form.

\subsection{Scalar field}

Let us construct momentum eigenstates for the scalar field $s=0$.
In this case the module \eqref{Phi_set_module} is
\begin{equation}
\Phi^{s=0}(Y,\tau)=\Phi(y\bar{y},\tau),\label{module_scalar}
\end{equation}
and the momentum operator \eqref{P_solution} reduces to
\begin{eqnarray}
P_{\alpha\dot{\alpha}}^{s=0} & = & \frac{i(\varsigma+3/2)}{(\varsigma+1)(\varsigma+2)}\partial_{\alpha}\bar{\partial}_{\dot{\alpha}}+\frac{i}{(\varsigma+1/2)}y_{\alpha}\bar{y}_{\dot{\alpha}}\frac{\partial^{2}}{\partial\tau^{2}}.
\end{eqnarray}
We have to solve an equation
\begin{equation}
P_{\alpha\dot{\beta}}\Phi_{p}(Y,\tau)=p_{\alpha\dot{\beta}}\Phi_{p}(Y,\tau)\label{Pf=00003Dpf}
\end{equation}
with some momentum $p_{\alpha\dot{\beta}}$, $p^{2}=m^{2}$.

A natural Ansatz is
\begin{equation}
\Phi_{p}(Y,\tau)=\Phi_{p}(-ip_{\alpha\dot{\alpha}}y^{\alpha}\bar{y}^{\dot{\alpha}})e^{\pm im\tau},
\end{equation}
where $\tau$-dependence gets fixed by the general solution \eqref{gen_sol}
and $p_{\alpha\dot{\alpha}}y^{\alpha}\bar{y}^{\dot{\alpha}}$ is the
only available Lorentz-invariant combination involving $Y$.

Using that
\begin{equation}
\partial_{\alpha}\bar{\partial}_{\dot{\alpha}}f(z_{\beta\dot{\beta}}y^{\beta}\bar{y}^{\dot{\beta}})=z_{\alpha\dot{\alpha}}(\varsigma+1)f'-z^{2}y_{\alpha}\bar{y}_{\dot{\alpha}}f'',
\end{equation}
where the prime means the derivative with respect to the entire argument
of $f$, one can rewrite \eqref{Pf=00003Dpf} as an ODE with respect
to the variable $u=-ip_{\alpha\dot{\alpha}}y^{\alpha}\bar{y}^{\dot{\alpha}}$
\begin{equation}
u\Phi''(u)+(\frac{3}{2}-u)\Phi'(u)-2\Phi(u)=0.\label{Kummer}
\end{equation}
This arises from the terms in \eqref{Pf=00003Dpf}, proportional to
$p_{\alpha\dot{\beta}}$. Strictly speaking, there is one more ODE
coming from \eqref{Pf=00003Dpf}, which is generated by terms proportional
to $y_{\alpha}\bar{y}_{\dot{\alpha}}$, but it represents a differential
consequence of \eqref{Kummer}.

\eqref{Kummer} is the Kummer's equation. Its solution regular at
$u=0$ is the confluent hypergeometric function
\begin{equation}
\Phi(u)={}_{1}F_{1}(2;\frac{3}{2};u).
\end{equation}
Thus, momentum-$p_{\alpha\dot{\beta}}$ eigenstate of the scalar field
is
\begin{equation}
\Phi_{p}(Y,\tau)={}_{1}F_{1}(2;\frac{3}{2};-ipy\bar{y})e^{\pm im\tau}.
\end{equation}

\subsection{Massless fields}

For a massless spin-$s$ field the module is \eqref{Phi_module_massless}.
It contains two $\pm s$ helicities and for both of them the momentum
operator reduces to
\begin{eqnarray}
P_{\alpha\dot{\beta}}^{m=0} & = & \frac{i(\varsigma+3/2)}{(\varsigma+s+1)(\varsigma-s+2)}\partial_{\alpha}\bar{\partial}_{\dot{\beta}}.
\end{eqnarray}
Introducing a polarization vector $\varepsilon_{\alpha\dot{\beta}}$,
orthogonal to the null momentum $p_{\alpha\dot{\beta}}$, $p^{2}=0$,
\begin{equation}
\varepsilon_{\alpha\dot{\beta}}p^{\alpha\dot{\beta}}=0,
\end{equation}
we choose following Ansätze for negative and positive helicites
\begin{equation}
\Phi_{p,\varepsilon}^{-}(Y,\tau)=(i\varepsilon_{\alpha\dot{\beta}}p_{\alpha}{}^{\dot{\beta}}y^{\alpha}y^{\alpha})^{s}\Psi(-ipy\bar{y})e^{\pm i\mu\tau},
\end{equation}
\begin{equation}
\Phi_{p,\varepsilon}^{+}(Y,\tau)=(i\varepsilon_{\beta\dot{\alpha}}p^{\beta}{}_{\dot{\alpha}}\bar{y}^{\dot{\alpha}}\bar{y}^{\dot{\alpha}})^{s}\Psi(-ipy\bar{y})e^{\pm i\mu\tau}.
\end{equation}
Here we made use of a $\mu$-deformed realization of $P^{2}$ \eqref{P^2_mu}.
Then for $\Psi$ one gets, analogously to the scalar field case, the
following Kummer's equation
\begin{equation}
u\Psi''(u)+(\frac{3}{2}+s-u)\Psi'(u)-2\Psi(u)=0
\end{equation}
whose regular at $u=0$ solution is
\begin{equation}
\Psi(u)={}_{1}F_{1}(2;\frac{3}{2}+s;u).
\end{equation}

\section{Conclusion\label{SEC_conclusions}}

In the paper we proposed a new way of implementing bosonic UIR of
$4d$ Poincaré group. We presented them as bunches of scalar fields
on $5d$ space with coordinates $\{Y^{A},\tau\}$ and found appropriate
realizations for Poincaré generators. These realizations possess some
distinguishing features: the mass operator $P^{2}$ is independent
of spinor coordinates $Y$, so that equations of motion become ODE
in a Lorentz-invariant proper time $\tau$ and follow from a simple
manifestly Poincaré-invariant action. Thus, our construction demonstrates
a separation of variables: e.o.m. governs the evolution in $\tau$,
while $Y$ parameterize the space of Cauchy data. The translation
generators are local differential operators in $\tau$, but non-local
in $Y$, hence $\tau$-evolution is local, while translations act
non-locally on the Cauchy hypersurface spanned by $Y$.

The simple form of e.o.m. allowed us to write down their general solutions.
Those contain two branches, corresponding to different sign-dependence
on $\tau$, similarly to positive and negative energy branches in
the standard QFT approach. We found a Poincaré-invariant inner product,
which is positive-definite for one of the branches.

For massless fields we modified the mass operator by introducing an
IR-regulator. This allowed us to have bounded in $\tau$ solutions
and the split into two branches. This modification is manifestly Poincaré-invariant
and is possible due to the large ambiguity in the form of the mass
operator, caused by the separation of variables. Our construction
is non-gauge, as we work directly with helicity-expanded fields: the
bunch of scalar fields mentioned before represents a bunch of helicities
of a spin-$s$ representation, connected by Poincaré transformations.
On the zero-mass shell $\pm s$-helicity components form closed subrepresentations,
so 'gauge-fixing' reduces to direct putting all intermediate-helicity
components to zero.

We also found the momentum-eigenstate solutions for the simplest cases
of a scalar field and massless fields. They have the form of the confluent
hypergeometric functions.

The construction, proposed in the paper, poses many problems for further
research. One of the most urgent is to develop appropriate canonical
structures and to define an analogue of the canonical quantization
procedure, regarding that some necessary elements are already presented
(a classical action, distinguished in a Lorentz-invariant way coordinate
$\tau$ that governs the evolution, two branches of classical solutions
etc). Other interesting directions include considering fermionic and
infinite-spin representations as well as supersymmetric extensions,
generalizations to (A)dS backgrounds and, the most important, introducing
interactions. The problem of interactions, in its turn, immediately
rise many questions: can one formulate a systematic procedure of looking
for Poincaré-invariant vertices? what happens to the separation of
$\tau$ and $Y$ variables at the nonlinear level? how does the $Y$-nonlocality
of Poincaré transformations affect the perturbative analysis? One
may hope that answering these questions will provide us with new powerful
formalism for studying higher-spin theories.

\section*{Acknowledgments}

The research was supported by the Alexander von Humboldt Foundation.

\end{document}